\documentclass[aps,prb,twocolumn,superscriptaddress]{revtex4}

\usepackage{graphicx}
\usepackage{dcolumn}	
\usepackage{bm}			
\usepackage{here}
\usepackage{color}
\usepackage{ulem}
\begin{document}


\title{Charge Order and Fluctuations in Bi$_2$Sr$_{2-x}$La$_x$CuO$_{6+\delta}$ Revealed by $^{63,65}$Cu-Nuclear Magnetic Resonance}


\author{Shinji Kawasaki}
\affiliation{Department of Physics, Okayama University, Okayama 700-8530, Japan}

\author{Madoka Ito}
\affiliation{Department of Physics, Okayama University, Okayama 700-8530, Japan}

\author{Dai Kamijima}
\affiliation{Department of Physics, Okayama University, Okayama 700-8530, Japan}

\author{Chengtian Lin}
\affiliation{Max-Planck-Institut fur Festkorperforschung, Heisenbergstrasse 1, D-70569 Stuttgart, Germany}

\author{Guo-qing Zheng}
\affiliation{Department of Physics, Okayama University, Okayama 700-8530, Japan}


\begin{abstract}

The discovery of a magnetic-field-induced charge-density-wave (CDW) order in the pseudogap state via nuclear magnetic resonance (NMR) studies has highlighted the importance of ``charge'' in the physics of high transition-temperature ($T_{\rm c}$) superconductivity in copper oxides (cuprates). Herein, after briefly reviewing the progress achieved in the last few years, we report new results of $^{63,65}$Cu-NMR measurements on the CDW order and its fluctuation in the single-layered cuprate Bi$_2$Sr$_{2-x}$La$_x$CuO$_{6+\delta}$. The NMR spectrum under both in- and out-of-plane magnetic fields above $ H $ = 10 T indicates that the CDW replaces the antiferromagnetic order before superconductivity appears, but disappears before superconductivity is optimized. We found that the CDW onset temperature $T_{\rm CDW}$ scales with the pseudogap temperature $T^{\rm *}$. Comparison between $^{63}$Cu and $^{65}$Cu NMR indicates that the spin-lattice relaxation process is dominated by charge fluctuations in the doping regions where the CDW appears as well as at the pseudogap end point ($T^*$ = 0). These results suggest that charge orders and fluctuations exist in multiple doping regions and over a quite wide temperature range.

\end{abstract}



    \maketitle
    
    \section{Introduction}
    Since the discovery of superconductivity in La$_{2-x}$Ba$_x$CuO$_4$  in 1986 \cite{Bednorz}, copper oxide (cuprate) has been thoroughly investigated towards a realization of room-temperature superconductivity \cite{Uchida}. One of the remarkable features of cuprate superconductivity is its Cooper pair symmetry, which has been established to be an anisotropic $d$-wave gap with spin singlet \cite{Tsuei}. This is a manifestation of strong electron correlations in contrast to the electron phonon interaction in conventional superconductors \cite{BCS}. Unfortunately, the superconducting transition temperature ($T_{\rm c}$) did not increase after the discovery of $T_{\rm c}$ = 134 K in HgBa$_2$Ca$_2$Cu$_3$O$_{8+\delta}$ in 1993 \cite{Schilling}. The mechanism of superconductivity remains unclear as well, although it is well accepted that it is tightly related to Mott insulating phase  \cite{PALee}. 
    
    Among others, the ``pseudogap'' phenomenon is an unsettled issue \cite{Timusk}, first found by nuclear magnetic resonance (NMR) experiments \cite{Yasuoka,Alloul}. The nuclear spin-lattice relaxation rate divided by the temperature $1/T_1T$ and Knight shift ($K$) decreases below a certain temperature (hereafter referred as $T^{\rm *}$). This indicates that the density of states (DOS) is reduced below $T^{\rm *}$ which is above an antiferromagnetic transition or $T_{\rm c}$ \cite{Yasuoka,Alloul,ZhengPRL,KawasakiPRL}.  The angle-resolved photoemission spectroscopy (ARPES) experiments have shown that the pseudogap opens around ($\pi$, 0) region in the momentum space below $T^{\rm *}$ above $T_{\rm c}$, whose momentum dependence closely resembles that of the $d_{x^2-y^2}$ gap observed in the superconducting state \cite{Marshall,Ding,Loeser}.  Thus, the focus has been on the relationship between the pseudogap and the superconducting gap \cite{Norman}. Regarding this, we have performed NMR measurements on single-layered Bi$_2$Sr$_{2-x}$La$_x$CuO$_{6+\delta}$ (Bi2201) superconductors after suppressing superconductivity by strong magnetic field and revealed that the two gaps are coexisting matter \cite{ZhengPRL,KawasakiPRL}. However, the cause for the DOS decrease below $T^{\rm *}$ is still unclear.
    
    Combining ARPES, polar Kerr effect, and time-resolved reflectivity experiments in Bi2201, it was argued that $T^*$ is associated with a phase transition to a symmetry-broken state \cite{ZXShen}. On the other hand, previous NMR experiments found no internal magnetic field due to magnetic order \cite{KawasakiPRL,Keller,Mounce} nor $C_4$ symmetry breaking below $T^{\rm *}$ \cite{Crocker}. Although a unified understanding of the origin of the pseudogap has not been reached, some clues have been obtained. In La-based cuprates, at hole concentration $p$ = 1/8, neutron scattering measurements have found a ``stripe'' phase due to the spin/charge order that competes with superconductivity \cite{Tranquada,Koike,HimedaOgata}. Indeed, when a strong magnetic field is applied to YBa$_2$Cu$_3$O$_y$ (YBCO) to suppress superconductivity, a long range ordered  charge density wave (CDW) shows up \cite{WuNature}.  The field-induced CDW is not limited to YBCO, but also found in Bi2201 \cite{KawasakiNatComm}, although the relationship between CDW and superconductivity is quite different among the two classes of compounds. In the former case, the CDW arises only from the mixed state around $p$ = 1/8 \cite{WuNature}, but in the latter case it appears above $T_{\rm c}$ to take over antiferromagnetism, coexisting with superconductivity and disappearing before superconductivity reaches optimum \cite{KawasakiNatComm}. Furthermore, scanning tunneling microscopy (STM) of Bi$_2$Sr$_2$CaCu$_2$O$_{8+\delta}$ (Bi2212) revealed a modulation of the local DOS in the vortex cores \cite{Hoffman}, which was interpreted as a result of incipient charge order localized within the vortex halos \cite{Sachdev,KivelsonLee}. In that region, a pair density wave (PDW) order in which the order parameter varies periodically in space has been proposed \cite{PDWreview}. These results suggest that multiple orders are intertwined in the pseudogap region, and ``density wave'' has become a hot topic \cite{PALeePDW,Fradkin,Tohyama}.  
    
    The rest of this article is organized as follows. In Section 2, we review CDWs observed in cuprates mainly  from an NMR  point of view.  In Sections 3, we report new results of CDW order and fluctuations obtained through $^{63,65}$Cu-NMR  measurements on Bi2201. In section 4, we discuss the results in connection to other orders such as nematic order and possible
    quantum criticality associated with pseudgap end point, before summarizing in Section 5.

    \section{Review of CDWs in cuprates}

    The  discovery of a magnetic field-induced CDW order in YBCO \cite{WuNature} highlights the importance of ``charge''. Thus far, the CDW order has been confirmed only in YBCO \cite{WuNature,Nojiri,Changhighfield} and Bi2201 \cite{KawasakiNatComm}. We review the current status of the CDW studies in this section.
    \subsection{Magnetic field induced CDW order}

    In 2011, Wu $et$ $al$ discovered a magnetic-field-induced CDW order in underdoped YBCO [Fig. 1(a)] \cite {WuNature}. According to their results, a CDW appears inside the superconducting dome in a field perpendicular to the CuO$_2$ plane ($H$$\parallel$$c$)  above $H$ = 15 T, but not in a parallel field ($H$$\perp$$c$) \cite {WuNature}. The transition temperature ($T_ {\rm CDW}$) is the highest at approximately $p$ = 1/8, but lower than the zero-field $T_{\rm c}$  [$T_{\rm CDW}$ $\leq$ $T_{\rm c}$ ($H$ = 0)] \cite {WuVortex}. In contrast, as seen in Fig. 1(b), the onset field of the CDW ($H_{\rm CDW}$) at $T$ = 0 scales with the upper critical field $H_{\rm c2}$ and is the smallest at $p$ = 1/8  \cite{WuVortex}. Further, the fact that the superconductivity is inherently suppressed at $p$ = 1/8, where the CDW is likely to be induced, suggests that the CDW and superconductivity are competing orders \cite{WuNature}.  From these results, it was suggested that the CDW may develop owing to an overlap of the local charge density modulation at the vortex center \cite{Hoffman} in the mixed state \cite{WuVortex}.  
    
    Following the experiments on YBCO, high-field NMR experiments were performed in HgBa$_2$CuO$_{4+\delta}$ (Hg1201) \cite{Halperin}.  No evidence for field- or temperature-induced spin/charge order was found nor was there any correlation with the vortex or with the pseudogap. Therefore, it was concluded that field-induced CDWs are absent in Hg1201 \cite{Halperin}.
    
    In 2017, we reported high-field NMR measurements up to $H$ = 45 T in single-layered Bi2201 superconductors and discovered  CDWs  induced by an $in$-$plane$ magnetic field $H$$\perp$$c$ \cite{KawasakiNatComm}. As shown in Figs. 1(a)-(d), the doping dependence of the CDW and its relation to superconductivity are different in YBCO and Bi2201. The CDW of Bi2201 appears above $T_{\rm c}$ to replace the antiferromagnetic order.  Furthermore, although $T_ {\rm CDW}$ is of the same order in both compounds, $H_{\rm CDW}$ and $T_{\rm CDW}$ seem to be related to the doping level itself in Bi2201, but not to $H_{\rm c2}$ or superconductivity.  Thus, the CDW in Bi2201 is more likely to be induced near the antiferromagnetic order than at the specific hole concentration $p = 1/8$. 
    This difference is also observed in high-temperature CDW correlations. Next, we briefly review the normal state properties.

    \begin{figure}
        
        \begin{center}
            \includegraphics[width=0.9\linewidth]{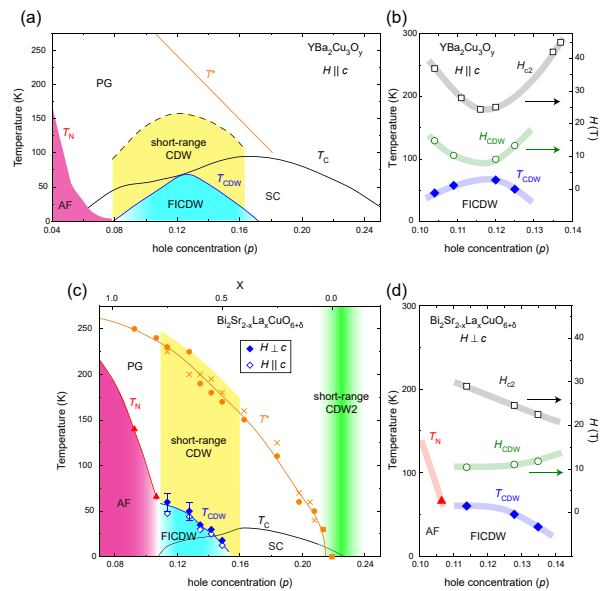}
            \caption{Doping dependence of short-range CDW/CDW2 \cite{CominBi2201,YYPeng,PengNatMat,Uchida}, field-induced CDW (FICDW), and its relation to upper critical field ($H_{\rm c2}$) and the onset field $H_{\rm CDW}$ for YBa$_2$Cu$_3$O$_{\rm y}$ (a), (b) and  Bi$_2$Sr$_{2-x}$La$_x$CuO$_{6+\delta}$ (c), (d). For both cases, $T_{\rm c}$ is the zero-field value.  The schematic phase diagram for YBa$_2$Cu$_3$O$_{\rm y}$ is  from the literatures \cite{Uchida,WuNature,WuVortex}.}
            \label{f1}
        \end{center}
    \end{figure}

    \subsection{CDW correlation length}
    After the aforenoted NMR study \cite{WuNature}, resonant X-ray scattering (RXS) experiments on YBCO at $H$ = 0 have revealed an in-plane short-range CDW [Fig. 1(a)] with wave vector ${\bf Q}$ = ($\sim$0.3, 0), (0, $\sim$0.3), and correlation lengths $\xi_{a, b}$$\sim$50$ {\AA}$ \cite{Ghiringhelli, ChangNatPhys}. The CDW appears below $T^{\rm *}$, but is suppressed below $T_{\rm c}$ \cite{Ghiringhelli, ChangNatPhys}. The onset temperature of the CDW is consistent with a peak in 1/$T_1T$ \cite{TakigawaT1T} and  has a similar doping dependence [dashed line in Fig. 1(a)]. Through $^{17}$O-NMR measurements, it was concluded that this CDW could be static \cite{WuONMR,Wang1248} in origin.  A similar short-range CDW was consistently found in Bi2201 \cite{CominBi2201,YYPeng}, Bi2212 \cite{SilvaNetoBi2212}, Hg1201 \cite{Tabis}, and even in the electron-doped cuprate Nd$_{2-x}$Ce$_x$CuO$_4$ \cite{Nd214}. Recently, in Bi2201, a re-entrant incommensurate charge order with substantially long  correlation lengths of 150-200 $\AA$  that persists up to temperatures of at least 250 K was found in the high doping region, above which the pseudogap is closed ($p$ $>$ 0.22) [short-range CDW2 in Fig. 1(c)] \cite{PengNatMat}.

    Further high field RXS experiments on YBCO revealed that the CDW correlations in the CuO chain and between the CuO$_2$ planes are enhanced by $H$$\parallel$$c$, leading to a transition to a three-dimensional long-range CDW \cite{Nojiri, Changhighfield}.  These observations  are consistent with the early discovery of a Fermi surface reconstruction by quantum oscillations \cite{Leyraud1} and  a  recent report of a thermodynamic phase transition in a high magnetic field \cite{LeBoeuf}.  

    \subsection{Similarity and difference between YBCO and Bi2201}
    As described above, the RXS experiments suggested that the short-range CDW is widely observed in cuprates, indicating that the CDW is a new key ingredient to understanding the cuprate physics. By contrast, the long-range CDW order has only been found in YBCO and Bi2201. It was suggested that the CDWs in YBCO are related to its local structure, and that the periodic oxygen deficiency in the CuO chain (Ortho II/VIII structure \cite{Andersen}) must be considered \cite{WuNature, WuONMR}. Particularly, even at zero magnetic field, it was found that a long-range three dimensional CDW is induced by uniaxial pressure parallel to the CuO chain, rather than perpendicular to it, suggesting a strong correlation between the CuO chain and the CDWs in YBCO \cite{YBCOfilm,YBCOstrain}. Such a correlation with lattice distortion is also true for the stripe order in the La-based cuprate \cite{Axe}.

    By contrast, Bi2201 is a single-CuO$_2$-layer compound without a CuO chain, and exhibits no structural transitions. The CDW correlation is perfectly two-dimensional and limited to the CuO$_2$ plane \cite{YYPeng}; further, the temperature at which short-range CDWs occur in Bi2201 coincides with $T^{\rm *}$ \cite{CominBi2201,YYPeng,ZhengPRL, KawasakiPRL}. Therefore, Bi2201 is of particular importance for the investigation of the CDWs. Moreover, the $T_ {\rm c}^{\rm max}$ of  Bi2201 (= 32 K) is the lowest among the $T_ {\rm c}^{\rm max}$ values of cuprates, which allows us to study the doping evolution of the ground states without superconductivity. It is also a significant advantage of Bi2201 that a single crystal can be obtained in a wide doping range, from an underdoped antiferromagnetic insulator to an overdoped metal above the pseudogap end point  [Fig. 1(c)]. 

    \section{New measurements and results in Bi2201}

    Hereafter, we report new results of $^{63,65}$Cu-NMR measurements of Bi2201 while reviewing the previous result \cite{KawasakiNatComm}. In the previous report, from the magnetic field and temperature dependence of the $^{63}$Cu-NMR spectrum, we found that, for the doping level at which superconductivity begins to emerge, a field-induced CDW dominates over the spin order. Such CDW disappears before the superconductivity becomes optimum \cite{KawasakiNatComm}.  Most importantly, we showed that $T_{\rm CDW}$ scales with $T^{\rm *}$ \cite{KawasakiNatComm}. As new experimental results, here we demonstrate that the CDW appears regardless of whether the magnetic field is applied perpendicular or parallel to the CuO$_2$ plane. The anisotropy of the CDW is considerably smaller than that of $H_{\rm c2}$.  Furthermore, from systematic $^{63}$Cu and $^{65}$Cu nuclear spin lattice relaxation time ($T_1$) measurements, we find that both spin and charge fluctuations are present in the normal state even above $T^{\rm *}$. The charge fluctuations become dominant in the doping regions where the field-induced CDW appears and at the pseudogap end point. Our results reveal that not only are the spin and charge orders of copper oxide intricately intertwined, but also their fluctuations, already in the normal state.

    \begin{table}
        \caption{List of single crystals of  Bi$_2$Sr$_{2-x}$La$_x$CuO$_{6+\delta}$. $T^{H \perp c}_{\rm CDW}$ and $T^{H \parallel c}_{\rm CDW}$ indicate the onset temperature of the field-induced CDW order above $H$ $\ge$ 13 T. Some data are from our previous papers \cite{ZhengPRL,KawasakiPRL,KawasakiNatComm}.  
        }
        \label{t1}
        \begin{center}
            \begin{tabular}{lclclclclclcl}
                \hline
                \multicolumn{1}{c}{La(x)}  & \multicolumn{1}{c}{$p$}  & \multicolumn{1}{c}{$T^{*}$(K)}  & \multicolumn{1}{c}{$T_{\rm N}$(K)} & \multicolumn{1}{c}{$T_{\rm c}$(K)}  & \multicolumn{1}{c}{$T^{H \perp c}_{\rm CDW}$(K)} & \multicolumn{1}{c}{$T^{H \parallel c}_{\rm CDW}$(K)}\\
                \hline
                \multicolumn{1}{c}{0.90} & \multicolumn{1}{c}{0.093} & \multicolumn{1}{c}{250} & \multicolumn{1}{c}{140}& \multicolumn{1}{c}{--}   & \multicolumn{1}{c}{--} & \multicolumn{1}{c}{--} \\
                \multicolumn{1}{c}{0.80} & \multicolumn{1}{c}{0.107} & \multicolumn{1}{c}{240} &\multicolumn{1}{c}{66} & \multicolumn{1}{c}{--}   & \multicolumn{1}{c}{--} & \multicolumn{1}{c}{--}  \\
                \multicolumn{1}{c}{0.75} & \multicolumn{1}{c}{0.114} &  \multicolumn{1}{c}{230}    & \multicolumn{1}{c}{--} & \multicolumn{1}{c}{12}   & \multicolumn{1}{c}{60} & \multicolumn{1}{c}{48}   \\
                \multicolumn{1}{c}{0.65} & \multicolumn{1}{c}{0.128} &  \multicolumn{1}{c}{210}  & \multicolumn{1}{c}{--} & \multicolumn{1}{c}{19}   & \multicolumn{1}{c}{50} & \multicolumn{1}{c}{44}   \\
                \multicolumn{1}{c}{0.60} & \multicolumn{1}{c}{0.135} &  \multicolumn{1}{c}{195}  & \multicolumn{1}{c}{--} & \multicolumn{1}{c}{20}   & \multicolumn{1}{c}{35} & \multicolumn{1}{c}{30}   \\
                \multicolumn{1}{c}{0.55} & \multicolumn{1}{c}{0.142} &  \multicolumn{1}{c}{190} & \multicolumn{1}{c}{--} & \multicolumn{1}{c}{23}   & \multicolumn{1}{c}{30} & \multicolumn{1}{c}{25}   \\
                \multicolumn{1}{c}{0.50} & \multicolumn{1}{c}{0.149} &  \multicolumn{1}{c}{175} & \multicolumn{1}{c}{--} & \multicolumn{1}{c}{23}   & \multicolumn{1}{c}{18} & \multicolumn{1}{c}{12.5}   \\
                \multicolumn{1}{c}{0.40} & \multicolumn{1}{c}{0.162} &  \multicolumn{1}{c}{160} & \multicolumn{1}{c}{--} & \multicolumn{1}{c}{32}   & \multicolumn{1}{c}{--} & \multicolumn{1}{c}{--}   \\
                \multicolumn{1}{c}{0.15} & \multicolumn{1}{c}{0.198} &  \multicolumn{1}{c}{65} & \multicolumn{1}{c}{--} & \multicolumn{1}{c}{22}   & \multicolumn{1}{c}{--} & \multicolumn{1}{c}{--}   \\
                \multicolumn{1}{c}{0.10} & \multicolumn{1}{c}{0.205} &  \multicolumn{1}{c}{60} & \multicolumn{1}{c}{--} & \multicolumn{1}{c}{19}   & \multicolumn{1}{c}{--} & \multicolumn{1}{c}{--}   \\
                \multicolumn{1}{c}{0.08} & \multicolumn{1}{c}{0.208} &  \multicolumn{1}{c}{45} & \multicolumn{1}{c}{--} & \multicolumn{1}{c}{14}   & \multicolumn{1}{c}{--} & \multicolumn{1}{c}{--}   \\
                \multicolumn{1}{c}{0.04} & \multicolumn{1}{c}{0.213} &  \multicolumn{1}{c}{30} & \multicolumn{1}{c}{--} & \multicolumn{1}{c}{10}   & \multicolumn{1}{c}{--} & \multicolumn{1}{c}{--}   \\
                \multicolumn{1}{c}{0.00} & \multicolumn{1}{c}{0.219} &  \multicolumn{1}{c}{--} & \multicolumn{1}{c}{--} & \multicolumn{1}{c}{8}   & \multicolumn{1}{c}{--} & \multicolumn{1}{c}{--}   \\
                \hline
            \end{tabular}
        \end{center}
    \end{table}



    \subsection{Samples}

    The single crystals of Bi$_2$Sr$_{2-x}$La$_x$CuO$_{6+\delta}$ listed in Table. 1 were grown by the traveling solvent floating zone method \cite{Liang,Peng}. The hole concentration ($p$) were estimated previously by the Hall coefficient \cite{Ono}. Small and thin single-crystal platelets, typically sized up to 2 mm-2 mm-0.1 mm, cleaved from an as-grown ingot, were used. The in-plane Cu-O bond direction ($a$ or $b$ axis) was determined by Laue reflection. $T_{\rm c}(H)$ is defined as the onset temperature of diamagnetism observed by ac-susceptibility measurement using NMR coil \cite{KawasakiNatComm}. 
    
    \subsection{NMR measurements}
    In Bi2201, there is only one Cu site in the single CuO$_2$ plane. For NMR measurements, the magnetic field is applied along the $c$ ($H$$\parallel$$c$) and the Cu-O bond direction ($H$$\perp$$c$), respectively. 
    The $^{63}$Cu-NMR spectra were taken by sweeping the rf frequency at a fixed field. The data above $H$ = 15 T were obtained by using the Hybrid magnet in the National High Magnetic Field Laboratory, Tallahassee, Florida \cite{KawasakiNatComm}.

    For $^{63,65}$Cu, the nuclear spin Hamiltonian is expressed as the sum of the Zeeman and nuclear quadrupole interaction terms, $\mathcal{H}$ = $\mathcal{H}_{\rm z} + \mathcal{H}_{\rm Q}$ = $-^{63,65}\gamma\hbar{\bf I}\cdot{\bf{H}_{0}} (1+K) + (h \nu_{\rm Q}/6)[3{I_z}^2-I(I+1)+\eta({I_x}^2-{I_y}^2)]$, where gyromagnetic ratio $^{63}\gamma$ = 11.285 MHz/T and $^{65}\gamma$ = 12.089 MHz/T, $K$ is the Knight shift, and $I$ = 3/2 is the $^{63,65}$Cu nuclear spin. The NQR frequency $\nu_{\rm Q}$ and the asymmetry parameter $\eta$ are defined as $\nu_{\rm Q}$  = $\frac{3eQV_{zz}}{2I(2I-1)h}$, $\eta$ $=$ $\frac{V_{xx} - V_{yy}}{V_{zz}}$, with $^{63}Q$ = $-0.211 \times 10^{-24}$ cm$^{2}$ and $^{65}Q$ = $-0.195 \times 10^{-24}$ cm$^{2}$  and $V_{\alpha \beta}$ being the nuclear quadrupole moment and the electric field gradient (EFG) tensor \cite{abragam}. The principal axis $z$ of the EFG is along the $c$ axis and $\eta$ =  0 \cite{ZhengNuQ}. Due to $\mathcal{H}_{\rm Q}$, as shown in Fig. 2(a) and 2(b), one obtains the NMR center line and the two satellite transition lines between $|m\rangle$ and $|m-1\rangle$, ($|m|$ = 3/2, 1/2), at $\nu_{m\leftrightarrow m-1} = ^{63,65}\gamma H_0(1+K) + (\nu_{\rm Q}/2)(3\cos^2\theta -1)(m-1/2)$ + second-order correction for $^{63}$Cu and $^{65}$Cu isotopes, respectively.  
    Here, $\theta$ is the angle between $\bf{H}$ and EFG. As shown in Fig. 2(c) and 2(d), in the antiferromagnetically ordered case below $T_{\rm N}$ and $T_{\rm CDW}$, an internal field $H_{\rm int}$ and spatial distribution of $\delta K$ and $\delta \nu_{\rm Q}$ due to antiferromagnetic \cite{KawasakiPRL} and CDW \cite{SrPt2As2,KawasakiNatComm} orders appear to modify the spectrum shape, respectively.

    The $^{63,65}$Cu-NMR $T_1$ were measured at the center peak ($H$$\parallel$$c$) by using a single saturating pulse. To obtain $T_1$, the recovery curve of nuclear magnetization was fitted by the theoretical function \cite{Narath}, $1-M(t)$/$M_{\rm 0}$ = 0.9exp($-6t/T_1$)+0.1exp($-t/T_1$) where $M_{\rm 0}$ and $M(t)$ were the nuclear magnetization in the thermal equilibrium and at a time $t$ after the saturating pulse. 
    In strongly correlated electron systems, usually, $T_1$ probes the spin fluctuation through the hyperfine coupling constant $A_{\bf q}$ as $^{63,65}$$(1/T_1^M$) $\propto$ $^{63,65}\gamma$$k_{\rm B}T\sum_{\bf q} \left | A_{\bf q} \right |^2 \chi_\perp^{\prime\prime}({\bf q}, \omega_0)/\omega_0$,  where $\omega_0$  is the NMR frequency \cite{MoriyaJPSJ} and ${\bf q}$ is a wave vector for a spin order. In the case that  $T_1$ probes the EFG fluctuation,  then it is dominated by $Q$ as $^{63,65}$($1/T_1^Q) \propto 3(2I+3)(^{63,65}Q^2)/[10(2I-1)I^2]$ \cite{Obata,SrPt2As2}. Hence, the origin of $T_1$ can be identified from the ratio $^{65}$(1/$T_1$)/$^{63}$(1/$T_1$), since $^{65}$(1/$T_1$)/$^{63}$(1/$T_1$) = ($^{65}\gamma$/$^{63}\gamma$)$^2$ = 1.15 when it is dominated by spin fluctuations and  $^{65}$(1/$T_1$)/$^{63}$(1/$T_1$) = ($^{65}Q$/$^{63}Q$)$^2$ = 0.85 when it is governed by charge fluctuations.

    \subsection{Field and temperature dependence of the Cu-NMR spectrum}
    \begin{figure}
        \begin{center}
            \includegraphics[width=0.9\linewidth]{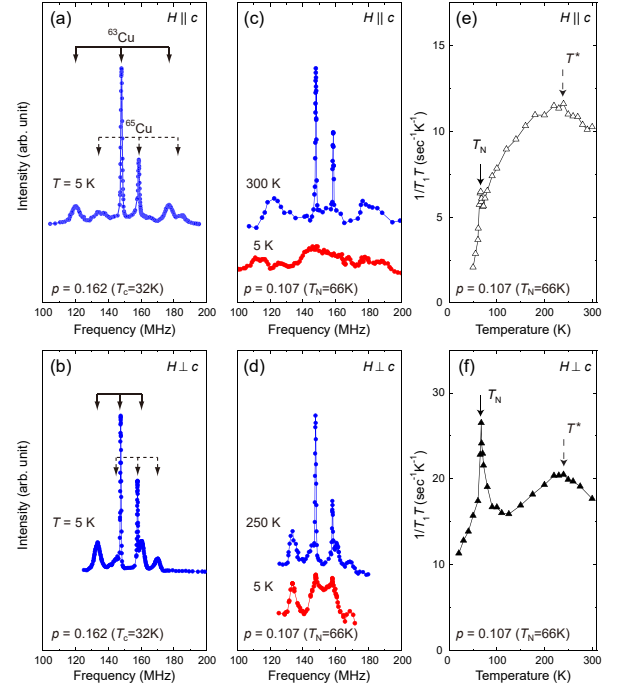}
            \caption{$^{63,65}$Cu-NMR spectrum of $p$ = 0.162 at $T$ = 5 K and that of $p$ = 0.107 above and below $T_{\rm N}$ at $H$ = 13 T obtained for $H$$\parallel$$c$  [(a) and (c)] and for $H$$\perp$$c$ [(b) and (d)], respectively. The lower satellite line (3/2$\leftrightarrow$1/2 transition) peak around 135 MHz was used for the detection of CDW for $H$$\perp$$c$. Temperature dependence of 1/$T_1T$ of $p$ = 0.107 in $H$ = 13 T obtained for $H$$\parallel$$c$ (e) and for $H$$\perp$$c$ (f). Solid and dashed arrows indicate $T_{\rm N}$ and $T^{\rm *}$. }
            \label{f1}
        \end{center}
    \end{figure}

    We observed the CDW order as a splitting of an NMR satellite peak due to in-plane EFG distribution \cite{KawasakiNatComm}. First, we show that this method is applicable even for the antiferromagnetic phase where $H_{\rm int}$ appears. 
    As shown in Fig. 2(c) and 2(d), for a strongly underdoped antiferromagnetic insulator ($p = 0.107$), the low-temperature spectrum changes its shape due to $H_{\rm int}$ below $T_{\rm N}$ = 66 K, where $1/T_1T$ exhibits a peak [Fig. 2(e) and 2(f)].  However, the in-plane $H_{\rm int}^{\perp}$ is significantly smaller than the out-of-plane $H_{\rm int}^{\parallel}$, \cite{KawasakiPRL} and thus, 
    the satellite peaks for $H$$\perp$$c$ is almost unchanged across $T_{\rm N}$. This allows us to verify the CDW even in the spin-ordered state.

    \begin{figure*}
        \centering
        \includegraphics[width=0.7\linewidth]{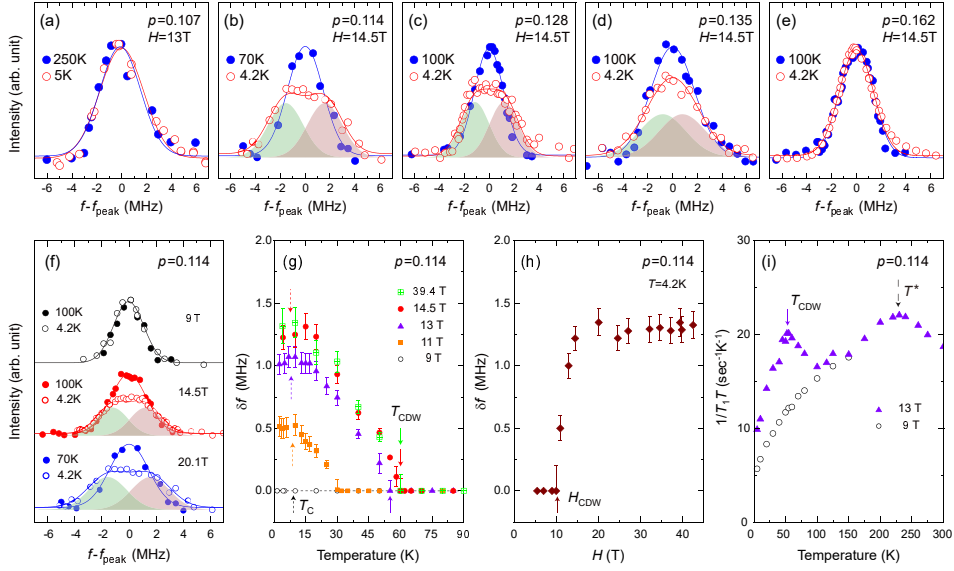}
        \caption{(a-e) NMR satellite  (3/2$\leftrightarrow$1/2 transition) lines for Bi$_2$Sr$_{2-x}$La$_x$CuO$_{6+\delta}$ with $p$ = 0.107, 0.114, 0.128, 0.135, and 0.162. The curves  for $p =$ 0.114, 0.128 and 0.135 at $T$ = 4.2 K are the sum of two Gaussian functions (shaded area).  $f_{\rm peak}$ is the peak frequency. 
            The intensity is normalized by the area of the spectrum. (f) Field and temperature dependence of the satellite line for $p =$ 0.114. (g) Temperature evolution of $\delta f$ for $p$ = 0.114 under various  fields.  The solid arrows indicate $T_{\rm CDW}$. The dotted arrows indicate $T_{\rm c}(H)$.   (h) Field evolution of the line splitting $\delta f$ for $p$ = 0.114 at $T$ = 4.2 K. The solid arrow indicates $H_{\rm CDW}$.  (i) The temperature dependence of $1/T_1T$ at different fields. The solid arrow indicate $T_{\rm CDW}$.  The dashed arrow indicates $T^{\rm *}$. Magnetic field is applied along the Cu-O bond direction ($H$$\perp$$c$). }
        \label{f1}
    \end{figure*}

    Figures 3(a)-(e) show the temperature dependence of the $^{63}$Cu-NMR satellite peak. The external magnetic field is $H$$\perp$$c$ = 13 and 14.5 T. First, no change was observed for $p$ = 0.162. For $p$ = 0.135, however, the line width increased at $T$ = 4.2 K. A splitting was also observed at $p$ = 0.128 and 0.114 but disappeared for $p$ = 0.107. As shown by the solid curves in the figure, the spectra for $p$ = 0.135, 0.128, and 0.114 can be reproduced by a sum of two Gaussian functions by formulating the splitting as $\pm\delta {f}$.
    
    To investigate the origin of the splitting, we focus on the results for $p$ = 0.114, in which case $\delta{f}$ is the largest. Figures 3(f) and 3(g) show the magnetic field and temperature 
    dependence of the satellite line and $\delta{f}(T)$ under various magnetic fields. $\delta{f}$ increases rapidly below $T_{\rm CDW}$ = 30 K, 55 K, and 60 K for $H$ = 11 T, 13 T, and above 14.5 T, respectively. Figure 3(h) shows the magnetic field dependence of $\delta{f}$ at $T$ = 4.2 K. $\delta{f}$ increases rapidly above $ H_{\rm CDW}$ = 10.4 T and becomes constant above $H$ = 14.5 T. Furthermore, as shown in Fig. 3(i), $1/T_1T$  shows a peak at $H$ = 13 T below which $\delta{f}$ starts to increase, although the peak is absent at $H$ = 9 T. These results indicate that a magnetic-field-induced phase transition occurs at $T_{\rm CDW}$ above $H_{\rm CDW}$ = 10.4 T.  

    Next, we quantitatively show that this phase transition is due to charge ordering, but not a spin order. Figures 4(a) and 4(b) respectively show the satellite and center peaks, obtained above and below $T_{\rm CDW}$ = 60 K, at $H$ = 14.5 T. 
    Figure 4(c) shows the temperature dependence of the integrated intensity of each peak multiplied by temperature ($I \times T$), which is constant above $T_{\rm c}$. 

    First, as shown in Fig. 2(d), in the case of the spin order, both the center and satellite lines become broader below $T_{\rm N}$ by the same amount $\sim$ 0.4 MHz due to $H^{\perp}_{\rm int}$. In contrast, as shown in Figs. 4(a) and 4(b), the satellite line becomes six times broader than the center line below $T_{\rm CDW}$. Second, it is also different from the spin/charge-stripe order observed at $p$ = 0.125 in the La-based cuprate \cite{Tranquada}. In Fig. 4 (c), a reduction of $I \times T$, which is a characteristic of the stripe order, the so-called ``wipe-out'' phenomenon, is not observed. Further, in the stripe order, the spatial distribution of the localized spins causes a very large $H_{\rm int}$ distribution \cite{AWHunt}, while the ``split'' of the spectrum below $T_{\rm CDW}$ shows evidence that the order parameter for the phase transition is spatially uniform. Finally, a magnetic-field-induced spin order was not found in our previous NMR experiments far above $ H_{\rm c2}$ at $p$ = 0.162 \cite{Mei}.

    These results indicate that the origin of the splitting of the spectra is not due to $H_{\rm int}$, but the spatial periodic distribution of the Knight shift $K \pm\delta{K}$ and the NQR frequency $\nu_{\rm Q}\pm\delta\nu$. Further, $\delta f_{\rm satellite}$ = 1.22 MHz is significantly larger than $\delta f_{\rm center}$ = 0.271 MHz, indicating that the origin is mainly $\nu_{\rm Q}$.  Here, the parameters $\delta{K}$ = 0.05$\pm$0.01 \% and $\delta{\nu}$ = 2.5$\pm$0.2 MHz can reproduce the two spectra simultaneously, as shown by the solid and dashed lines in Figs. 4(a) and 4(b) \cite{KawasakiNatComm}. In Bi2201, $\nu_{\rm Q}$ = 22.0+39.6$p$ was obtained \cite{KawasakiNatComm}, and from this, the spatial distribution of the hole concentration at the Cu site $\delta{p}$ = 0.06$\pm$0.01 was determined. Therefore, we conclude that the origin of $\delta{f}$ is a magnetic-field-induced uniform charge distribution ($\delta{p}$$\sim$0.06), i.e., the formation of the long-range CDW order below $T_{\rm CDW}$.

    \begin{figure}
        \begin{center}
            \includegraphics[width=0.98\linewidth]{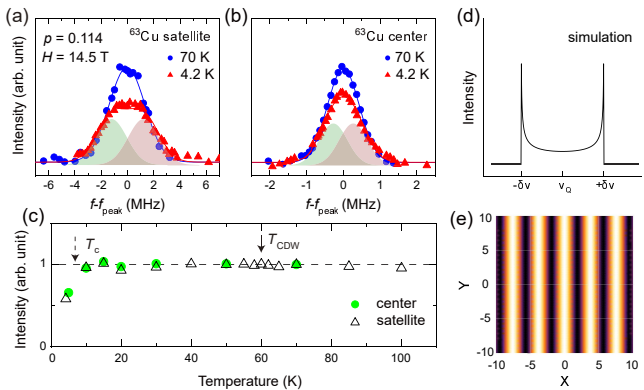}
            \caption{ Temperature dependence of satellite (a) and center (b) peaks of $p$ = 0.114 sample obtained at $H_\parallel$ = 14.5 T. The solid line is the fitting with a Gaussian function. The curve for $T$ = 4.2 K is the sum of two Gaussian functions. (c) Temperature dependence of NMR intensities ($I \times T$) for the center (solid circles) and satellite (open triangles) peaks. Dotted and solid arrows indicate $T_{\rm c}$ ($H$ = 14.5 T) and $T_{\rm CDW}$, respectively. The decrease below $T_{\rm c}$ is due to the Meissner effect. (d) Frequency distribution of Cu-NMR spectrum reflecting the incommensurate charge distribution in the CuO$_2$ plane. (e) Schematic view of the spatial distribution of charges $\delta{\nu}$$\cos$$({qX})$  below $T$ = $T_{\rm CDW}$ in the CuO$_2$ plane.   Yellow and black colors  indicate $+\delta{\nu}$ and $-\delta{\nu}$, respectively. $X$ and $Y$ indicate the positions of Cu along the Cu-O bond directions. }
            \label{f1}
        \end{center}
    \end{figure}

    \subsection{Possible CDW structure in the CuO$_2$ plane}
    The ``splitting'' of the satellite peaks can be understood by the one-dimensional (1D) incommensurate CDW order in the CuO$_2$ plane. For the 1D CDW, the NQR frequency is distributed spatially as $\nu$ = $\nu_{\rm Q} + \delta{\nu}$$\cos$$({qX})$ \cite{Blinc, SrPt2As2}. $X$ (= $a$ or $b$ axis) is the modulation direction and $2\delta{\nu}$ ($\propto{\delta{f}}$) is the CDW order parameter \cite {Blinc, SrPt2As2}. As a result, the spectrum exhibits a peak at $\nu$ = $\nu_{\rm Q}$$\pm$$\delta{\nu}$ [Fig. 4(d)] \cite{Blinc, SrPt2As2}, and the Gaussian function reproduces the spectrum below $T_{\rm CDW}$ if an appropriate line width is given.   Notably, although $T_{\rm CDW}$ shows similar values, the value of $\delta p$, that is, the amplitude of the CDW, is twice that for YBCO \cite{WuNature}. This is because Bi2201 has a single CuO$_2$ plane, and the CDW correlation is limited to this plane \cite{YYPeng}. Thus, it is considered that YBCO entails an effect of canceling out between its bi-layer CuO$_2$ planes \cite{Changhighfield}. In YBCO, the magnetic field enhances the CDW correlation in the CuO chain and the coupling between the CuO$_2$ planes, resulting in a three-dimensional CDW order \cite{Nojiri,Changhighfield}. By contrast, in Bi2201, the magnetic field yields a CDW \cite{CominBi2201,YYPeng} with a 1D long-range order only limited in the CuO$_2$ plane.

    The RXS experiment suggests a two-dimensional (2D) CDW with $\bf{Q}$ = (0.26, 0) and (0, 0.26) and $\xi$$\sim$20{\AA} below $T^{\rm *}$ when $H$ = 0 \cite{YYPeng, CominNatMat}. Furthermore, STM experiments on Bi2201 suggested a PDW with $\bf{q}_{\rm DW}$ = (0.25, 0) and (0, 0.25) on the surface at low temperatures \cite{DavisPNAS}. In either case, if  such a local CDW transforms  to a long-range order, the spatial distribution of the NQR frequency will be $\nu$ = $\nu_{\rm Q}$+$\delta\nu_X$$\cos$$(qX)$+$\delta\nu_Y$$\cos$$(qY)$ \cite{Blinc}. When the CDW amplitudes are equivalent to $\delta\nu_X$ = $\delta\nu_Y$, such CDWs yield $I(\nu){\sim}-{\rm ln}[(\nu-\nu_{\rm Q})/\delta\nu]/\delta\nu$, resulting in a logarithmic singularity at  $\delta\nu$ = 0 \cite{Blinc}. Therefore, such a long-range 2D CDW order is not consistent with our results. However, if the amplitude in each direction is very different, $\delta\nu_{X(Y)}\gg\delta\nu_{Y(X)}$ or there are  CDW  domains with modulations $\nu_X$ = $\nu_{\rm Q}$+$\delta\nu_X$$\cos$$(qX)$ and $\nu_Y$ = $\nu_{\rm Q}$+$\delta\nu_Y$$\cos$$(qY)$  in each domain, the NMR lineshape will be the same as that for the 1D CDW. Figure 4 (e) schematically shows the spatial modulation of the NQR frequency, $\delta{\nu}$$\cos$$({qX})$,  by assuming that a short-range CDW with $\bf{Q}$ = (0.26, 0) \cite{CominNatMat} becomes a long-range order under a magnetic field.

    \subsection{Relationship between CDW and superconductivity}

    To clarify the relationship between the CDW and superconductivity in more detail, we investigated the anisotropy of the CDW.  Figure 5(a) shows the doping dependence of the CDW order in $H$ $\parallel$ $c$ = 13 T determined by the temperature dependence of 1/$T_1T$ \cite{KawasakiNatComm}. Figure 5(b) shows the phase diagram of antiferromagnetism, CDW, and superconductivity in $H$ = 13 T. The anisotropy of the CDW is considerably smaller than that of the upper critical field for the superconductivity \cite{KawasakiNatComm}.  Therefore, vortices are not involved in the origin of CDWs. Rather, our results suggest that the CDW and superconductivity can coexist \cite{KawasakiNatComm}. This conclusion is supported by a more recent work for YBCO with $H$$\parallel$$c$ \cite{JulienPRL}.

    \begin{figure}
        \begin{center}
            \includegraphics[width=0.65\linewidth]{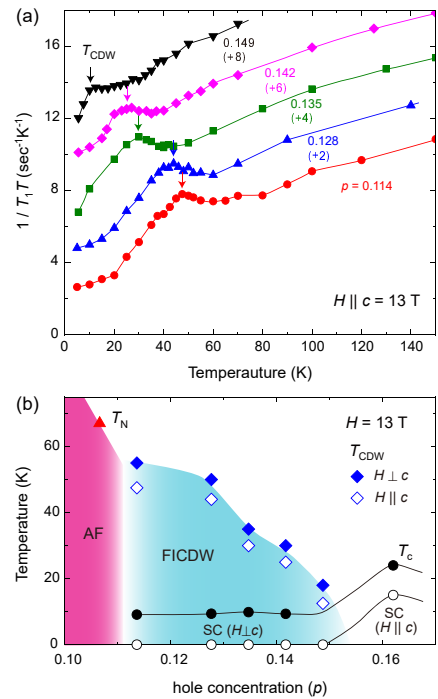}
            \caption{ (a) Temperature dependence of 1/$T_1T$ for $p$ = 0.114 to 0.149 obtained at $H$$\parallel$$c$ = 13 T. The data is offset for clarity. Arrows indicate $T_{\rm CDW}$. (b) Phase diagram of antiferromagnetism (AF), FICDW, and superconductivity (SC) at $H$ = 13 T. $T_{\rm c}$s are from ref \cite{KawasakiNatComm}.}  
            \label{f1}
        \end{center}
    \end{figure}

    \subsection{Relationship between CDW, antiferromagnetism, and pseudogap}
    In this subsection, we summarize the relationship between CDW and other phenomena.  First, $T_{\rm N}$ = 66 K for $p$ = 0.107 appears to be smoothly replaced by $T_{\rm CDW}$ = 60 K for $p$  = 0.114 where superconductivity appears. This indicates that CDW is more likely to be induced at the border of antiferromagnetism than at $p$ $\sim$ 1/8. Here, the pseudogap ground state of $p$ $\geq$ 0.114 is a metallic state with a finite density of states \cite{ZhengPRL,KawasakiPRL}. Therefore, the degree of freedom of charge is as important as that of spin in the development of high-temperature superconductivity. Second, the CDW is suppressed with increasing hole doping and disappears before the optimum $T_{\rm c}$  is reached at $p$ = 0.162. Most importantly, as shown in Fig. 6, we find that $T_{\rm CDW}$ and $T^{\rm *}$ have a positive correlation for the entire doping range. Because no static order is observed at $T^{\rm *}$ by NMR, Fig. 6 suggests that $T^{\rm *}$ is attributable to the CDW (charge) fluctuation as discussed in the next subsection. 
    Polar Kerr effect \cite{ZXShen} and optical conductivity measurements \cite{Hsieh} have suggested a symmetry breaking at $T^{\rm *}$. Note that, however, the timescales of these experiments are several orders of magnitude faster than NMR time scale. In view of this difference, these results may be consistent since $T^{\rm *}$ may appear as  a crossover in probes with a slower time windows. 

    \begin{figure}
        \begin{center}
            \includegraphics[width=0.72\linewidth]{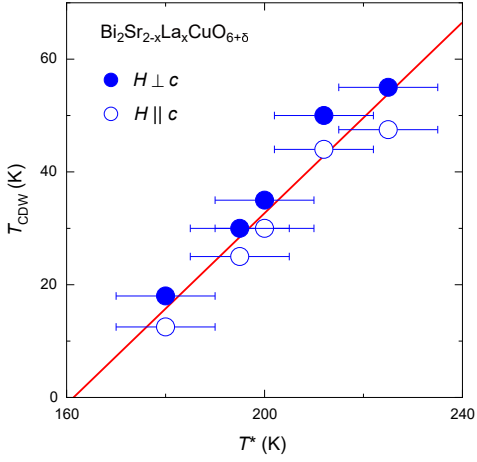}
            \caption{A plot of  $T_{\rm CDW}$ vs $T^{\rm *}$ at $H$ = 13 T. The straight line is a linear fitting to the data, which yields a slope of 0.9 $\pm$ 0.1. Error bars represent the uncertainty in defining $T^{\rm *}$. }
            \label{f1}
        \end{center}
    \end{figure}

    \subsection{CDW fluctuations}

    In this subsection, we report the results of the spin/charge fluctuations in the CuO$_2$ plane through the $^{63,65}$Cu NMR magnetic/quadrupolar relaxation. 
    Figures 7(a)-(c) show the recovery curves of $^{63}$Cu and $^{65}$Cu nuclear magnetization obtained at $T$ = 300 K for the antiferromagnetic insulator with $p$ = 0.107, CDW metal with $p$ = 0.114, and superconductor with $p$ = 0.162, respectively. A magnetic field $H$ = 13 T was applied along the $c$ axis. First, both $^{63,65}$Cu nuclear magnetization curves for $p$ = 0.107 and 0.162 are the same [$^{65}$(1/$T_1$)/$^{63}$(1/$T_1$) = 1], indicating that the relaxation process entails a mixture of spin and charge fluctuations  in the normal state, regardless of the antiferromagnetic insulator or superconductors. In contrast, we observed a longer $T_1$ for $^{65}$Cu than for $^{63}$Cu at $p$ = 0.114, which yields $^{65}$(1/$T_1$)/$^{63}$(1/$T_1$) = 0.85, i.e., charge fluctuations are dominant in the normal state of the CDW ordered phase.  
    
    Early NMR studies on polycrystalline YBCO \cite{KitaokaT1,ImaiT1} have suggested that the spin fluctuation was predominant and there was no charge fluctuation in the CuO$_2$ plane. 
    However, our results show that in addition to spin fluctuations, charge fluctuations are active in the CuO$_2$ plane.

    \begin{figure}
        \begin{center}
            \includegraphics[width=1\linewidth]{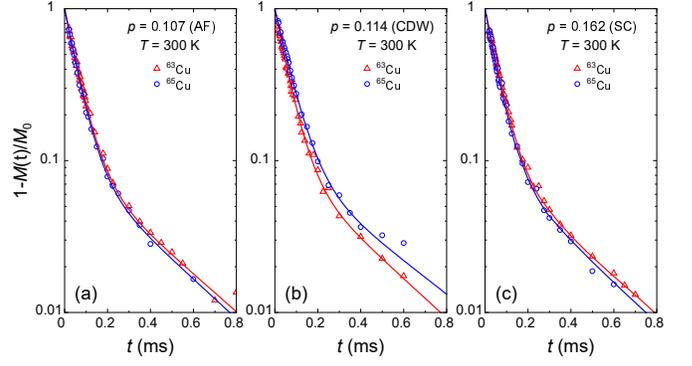}
            \caption{Typical recovery curves of the nuclear magnetization for $^{63}$Cu and $^{65}$Cu obtained at $T$ = 300 K for $p$ = 0.107 (a), 0.114 (b),  and 0.162 (c). Solid curves are fits to the theoretical formula, $1-M(t)$/$M_{\rm 0}$ = 0.9exp($-6t/T_1$)+0.1exp($-t/T_1$) \cite{Narath}.}
            \label{f1}
        \end{center}
    \end{figure}

    \begin{figure}
        \begin{center}
            \includegraphics[width=0.82\linewidth]{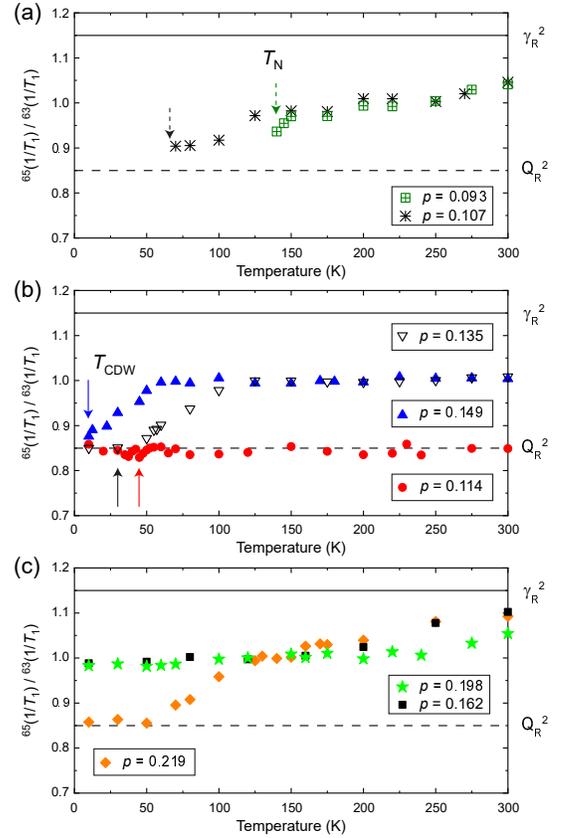}
            \caption{ Temperature dependence of the ratio $^{65}$(1/$T_1$)/$^{63}$(1/$T_1$) for strongly-underdoped antiferromagnets (a), underdoped CDW ordered superconductors (b), and overdoped superconductors (c) obtained at $H$ = 13 T parallel to the $c$ axis. Solid and dotted arrows indicate $T_{\rm CDW}$ and $T_{\rm N}$, respectively. Solid and dashed horizontal straight lines indicate the values expected for the magnetic relaxation process [$\gamma_{\rm R}^2$ ( = 1.15)]  and for the quadrupole relaxation process [${\rm Q}_{\rm R}^2$ ( = 0.85)], respectively.  }
            \label{f1}
        \end{center}
    \end{figure}

    We summarize the relationship between spin/charge order and fluctuations and pseudogap in Bi2201. Figures 8(a)-(c) show the temperature dependence of the ratio $^{65}$(1/$T_1$)/$^{63}$(1/$T_1$) for strongly underdoped antiferromagnets with $p$ = 0.093 and 0.107 [Fig. 8(a)], underdoped CDW-ordered superconductors with 0.114 $\leq$ $p$ $\leq$ 0.149 [Fig. 8(b)], and overdoped superconductors with 0.162 $\leq$ $p$ $\leq$ 0.219 [Fig. 8(c)].  Figure 9 shows the phase diagram for Bi2201, including the information on fluctuations in $H$$\parallel$$c$ = 13 T. $T_{\rm N}$, $T_{\rm CDW}$, $T_{\rm c}(H = 0)$, and $T^{*}$  with respect to La ($x$) and the hole concentration ($p$) are plotted with the magnitude of the $T_1$ ratio indicated by color. Spin and charge fluctuations are both present above $T_{\rm N}$ in the antiferromagnetic insulators ($p$ = 0.093 and 0.107). Notably, for $p$ = 0.114, where $T_{\rm CDW}$ is the highest, the charge fluctuation dominates over the entire temperature range.  With increasing doping, both the spin and charge fluctuations come into play again for $p$ = 0.135 and 149 superconductors with the charge fluctuations dominating below $T^{*}$. The charge fluctuations becomes less notable, but persist along with spin fluctuations as the CDW order disappears in the superconductors with $p$ = 0.162 and 0.198.

    Another surprising behavior was observed in the heavily overdoped superconductor with $p$ = 0.219, above which the pseudogap closes ($T^{\rm *}$ = 0). A mixture of spin and charge fluctuations is still visible at $p$ = 0.219, but spin fluctuation is dominant above $T$ = 250 K. As the temperature is lowered, the spin fluctuation is reduced and the charge fluctuation becomes dominant below $T$ = 50 K, although the CDW order is $absent$ at $p$ = 0.219.  We note that a stripe ordering below $T$ = 80 K has been found in the La-based cuprate at $p$ = 0.21, which was suggested by neutron scattering experiment to be  derived from spin fluctuations \cite{Koike}. However, spin fluctuation is not noticeable in Bi2201. Therefore, our observation at the pseudogap end point in Bi2201 is not related to the stripe ordering.

    \begin{figure}
        \begin{center}
            \includegraphics[width=0.95\linewidth]{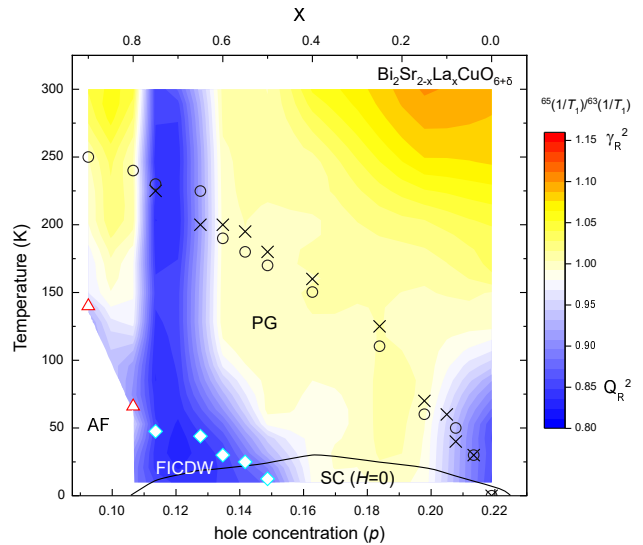}
            \caption{ Carrier concentration dependence of pseudogap (PG), antiferromagnetism (AF), field-induced CDW (FICDW), and superconductivity [SC ($H$ = 0)] for Bi$_2$Sr$_{2-x}$La$_x$CuO$_{6+\delta}$. The color coding (bar on right) refers to the ratio $^{65}$(1/$T_1$)/$^{63}$(1/$T_1$). Red, blue, and yellow shadings indicate the spin, charge, and mixed fluctuations, respectively. Open triangles, diamonds, and circles (crosses) indicate $T_{\rm N}$, $T_{\rm CDW}$, and $T^{\rm *}$, respectively.  Solid curve indicates $T_{\rm c}$ at $H$ = 0. $\gamma_{\rm R}^2$ = 1.15 and ${\rm Q}_{\rm R}^2$ = 0.85 are indicated on the bar, respectively.}
            \label{f1}
        \end{center}
    \end{figure}
    
    \section{Discussion}

    Our results revealed that charge fluctuations coexist with spin fluctuations over the entire doping range from high temperatures above $T^{\rm *}$ to low temperatures where superconductivity occurs. This supports the suggestion that the $T^{\rm *}$ is associated with the charge fluctuation. In contrast, at high temperatures and heavily overdoped regions away from the pseudogap, the spin fluctuation dominates over the  charge fluctuation (orange region in Fig. 9). 
    
    In two regions near the endpoints of the superconducting dome, the charge fluctuation dominates. The first one is the region where field-induced CDWs occur. The charge fluctuation dominates the spin fluctuation at the border of antiferromagnetism. This is in good agreement with the replacement of $T_{\rm CDW}$ with $T_{\rm N}$.  The other is at the pseudogap end point around $p$ $\sim$ 0.22. These results are qualitatively consistent with X-ray measurements at $H$ = 0 [Fig. 1(c)] \cite{CominBi2201,YYPeng,PengNatMat,CominNatMat}. The origin of the charge fluctuation in the two different regions, however, may be different. The former is a precursor to the CDW order. In the latter case, it is probably due to quantum critical fluctuations around the pseudogap endpoint.   In fact, a growth of nematic susceptibility or specific heat was observed in several compounds at the pseudogap end point \cite{Ishida,Girod,Lizaire}.

    Finally, a nematic order was recently reported in the pseudogap regime by torque magnetometry on YBa$_2$Cu$_3$O$_y$ \cite{Matsuda} and by  $^{17}$O-NMR experiments on YBa$_2$Cu$_4$O$_8$ \cite{Wang1248}. 
    The relationship and interplay between CDW orders/fluctuations and nematicity await future studies. Also, experiments at different time/length scales and magnetic-field dependence are important to further investigate the role of charge.

    \section{Summary}
    We reviewed the CDW orders in  Bi$_2$Sr$_{2-x}$La$_x$CuO$_{6+\delta}$ with comparison to YBCO and Bi2212, and presented our new NMR results on the CDW order and fluctuations.
    We clarified the relationship between spin/charge dynamics/orders and pseudogap.  We find that the CDW appears, regardless of  the magnetic field parallel or perpendicular  to the CuO$_2$ plane. However, the anisotropy of the CDW is considerably smaller than that of $H_{\rm c2}$. The CDW takes over the spin order in the region $p$ $\geq$ 0.114  and coexist with superconductivity, but disappears before the optimal doping level is reached.  We find $T_{\rm CDW}$ scales with $T^{\rm *}$. The charge fluctuations coexist with spin fluctuations in the CuO$_2$ plane in the entire doping and temperature regions, but  are dominant in the two end regions of the superconducting dome. In the first region with 0.114 $\leq$ $p$ $\leq$ 0.149, the field-induced CDW order appears, while in the second region where the pseudogap ends, charge fluctuations show up. These results demonstrate the usefulness of NMR in research on both static and dynamic properties, and hopefully will provide some hints for the future research in cuprates.

        We would like to thank Z. Li, M. Kitahashi, P. L. Kuhns, and A. P. Reyes for contribution in the early part of this work, and T. Tohyama for discussion.
        This work was supported in part by research grants from MEXT (Nos. JP19K03747 and JP19H00657).


    


\begin{thebibliography}{9}
        \bibitem{Bednorz}J. G. Bednorz and K. A. M\"{u}ller, Z. Phys. {\bf 64}, 189 (1986).
        \bibitem{Uchida}B. Keimer $et$ $al$., Nature {\bf 518}, 179 (2015). 
        \bibitem{Tsuei}
        C. C. Tsuei and J. R. Kirtley, Rev. Mod. Phys. {\bf 72}, 969 (2000).
        
        \bibitem{BCS}
        J. Bardeen, L. N. Cooper, and J. R. Schrieffer, Phys. Rev. {\bf 108}, 1175 (1957).
        
        \bibitem{Schilling}A. Schilling $et$ $al$., Nature {\bf 363}, 56 (1993).
        \bibitem{PALee}
        P. A. Lee, N. Nagaosa, and X. -G. Wen, Rev. Mod. Phys. {\bf 78}, 17 (2006).
        \bibitem{Timusk}T. Timusk and B. Statt, Rep. Prog. Phys. {\bf 62}, 61 (1999).
       
        \bibitem{Yasuoka}H. Yasuoka, T. Imai, and T. Shimizu, $Strong$ $Correlation$ $and$ $Superconductivity$, ed. H. Fukuyama, S. Maekawa, and A. P. Malozemoff (Springer-Verlag, New York, 1989) p. 254.
        \bibitem{Alloul}H. Alloul, T. Ohno, and P. Mendels, Phys. Rev. Lett. {\bf 63}, 1700 (1989).
        
        \bibitem{ZhengPRL}G.-q. Zheng $et$ $al$., Phys. Rev. Lett. {\bf 94}, 047006 (2005).
        \bibitem{KawasakiPRL}Shinji Kawasaki $et$ $al$., Phys. Rev. Lett. {\bf 105}, 137002 (2010).
        
        \bibitem{Marshall}
        D. S. Marshall $et$ $al$., Phys. Rev. Lett. {\bf 76}, 4841 (1996).
        \bibitem{Ding}
        H. Ding $et$ $al$., Nature {\bf 382}, 51 (1996).
        \bibitem{Loeser}
        A. G. Loeser $et$ $al$., Science {\bf 273}, 325 (1996). 
        
        \bibitem{Norman}
        M. R. Norman $et$ $al$., Nature {\bf 392}, 157 (1998).
        
        
        \bibitem{ZXShen}R. -H. He $et$ $al$., Science {\bf 331}, 1579 (2011).
        
        \bibitem{Keller}
        S. Stra\"{s}sle $et$ $al$., Phys. Rev. Lett. {\bf 106}, 097003 (2011).
        
        \bibitem{Mounce}
        A. M. Mounce $et$ $al$., Phys. Rev. Lett. {\bf 111}, 187003 (2013).
        
        \bibitem{Crocker}
        J. Crocker $et$ $al$., Phys. Rev. B {\bf 84}, 224502 (2011).
        
        \bibitem{Tranquada}J. M. Tranquada $et$ $al$., Nature {\bf 375}, 561 (1995).
        
        \bibitem{Koike}
        Y. Koike $et$ $al$.,  Physica C {\bf 364-365}, 562 (2001).
        
        \bibitem{HimedaOgata}
        A. Himeda, T. Kato, and M. Ogata, Phys. Rev. Lett. {\bf 88}, 117001 (2002).
        
        \bibitem{WuNature}T. Wu $et$ $al$., Nature {\bf 477}, 191 (2011).
        
        \bibitem{KawasakiNatComm}S. Kawasaki $et$ $al$., Nat. Commun. {\bf 8}, 1267 (2017).
        
        \bibitem{Hoffman}J. E. Hoffman $et$ $al$.,  Science {\bf 295}, 466 (2002).
        \bibitem{Sachdev}
        T. Zhang, E. Demler, and S. Sachdev, Phys. Rev. B {\bf 66}, 094501 (2002).
        
        \bibitem{KivelsonLee}
        S. A. Kivelson $et$ $al$., Phys. Rev. B {\bf 66}, 144516 (2002).
        
        
        
        \bibitem{PDWreview}
        Daniel F. Agterberg $et$ $al$.,  Annu. Rev. Condens. Matter Phys. {\bf 11}, 231 (2020).
        
        
        \bibitem{Tohyama} T. Tohyama, Jpn. J. Appl. Phys. {\bf 51}, 010004 (2011).
        \bibitem{PALeePDW}
        Patrick A. Lee, Phys. Rev. X {\bf 4}, 031017 (2014).
        
        
        \bibitem{Fradkin}
        E. Fradkin, S. A. Kivelson, and J. M. Tranquada, Rev. Mod. Phys. {\bf 87}, 457 (2015).
       
        
        \bibitem{Nojiri}S. Gerber $et$ $al$.,  Science {\bf 350}, 949 (2015).
        \bibitem{Changhighfield}
        J. Chang $et$ $al$., Nat. Commun. {\bf 7}, 11494 (2016). 
        
        \bibitem{WuVortex}T. Wu $et$ $al$.,  Nat. Commun. {\bf 4}, 2113 (2013).
        \bibitem{Halperin} J. A. Lee $et$ $al$.,  New. J. Phys. {\bf 19}, 033024 (2017).
        
               
        
        \bibitem{Ghiringhelli}G. Ghiringhelli $et$ $al$., Science {\bf 337}, 821 (2012).
        \bibitem{ChangNatPhys}J. Chang $et$ $al$., Nature Phys. {\bf 8}, 871 (2012).
        
        \bibitem{TakigawaT1T}
        M. Takigawa $et$ $al$., Phys. Rev. B {\bf 43}, 247 (1991).
        
        \bibitem{WuONMR}T. Wu $et$ $al$., Nat. Commun. {\bf 6}, 6438 (2015).
        \bibitem{Wang1248}
        W. Wang $et$ $al$.,  Sci. China-Phys. Mech. Astron. {\bf 64}, 237413 (2021).
        
        
        
        
        \bibitem{CominBi2201}
        R. Comin $et$ $al$., Science {\bf 343}, 390 (2014).
        
        \bibitem{YYPeng}Y. Y. Peng $et$ $al$., Phys. Rev. B {\bf 94}, 184511 (2016).
        
        
        \bibitem{SilvaNetoBi2212}
        E. H. da Silva Neto $et$ $al$., Nat. Commun. {\bf 5}, 5875 (2014).
        
        \bibitem{Tabis}W. Tabis $et$ $al$., Nat. Commun. {\bf 5}, 5875 (2014).
        
        
        \bibitem{Nd214} 
        E. H. da Silva Neto $et$ $al$., Science {\bf 347}, 282 (2015).
        
        \bibitem{PengNatMat}
        Y. Y. Peng $et$ $al$.,  Nature Mater. {\bf 17}, 697 (2018). 
        
        
        \bibitem{Leyraud1}
        N. Doiron-Leyraud $et$ $al$.,  Nature {\bf 447}, 565 (2007).
        
        \bibitem{LeBoeuf}
        D. LeBoeuf $et$ $al$., Nature Phys. {\bf 9}, 79 (2013).
        
        \bibitem{Andersen}N. H. Andersen $et$ $al$., Physica C {\bf 317-318}, 259 (1999).
                
        \bibitem{YBCOfilm}
        M. Bluschke $et$ $al$., Nat. Commun. {\bf 9}, 2978 (2018). 

        \bibitem{YBCOstrain}
        H.-H. Kim $et$ $al$., Science {\bf 362}, 1040 (2018).
        
        \bibitem{Axe}
        J. D. Axe $et$ $al$., Phys. Rev. Lett. {\bf 62}, 2751 (1989).
        
        
        \bibitem{Peng}
        J. B. Peng and C. T. Lin, J. Supercond. Nov. Magn. {\bf 23}, 591 (2010).
        
        \bibitem{Liang}
        B. Liang and C. T. Lin, J. Cryst. Growth {\bf 267}, 510 (2004).
        \bibitem{Ono}
        S. Ono $et$ $al$., Phys. Rev. Lett. {\bf 85}, 638 (2000). 
        
        \bibitem{abragam}A. Abragam, The principles of nuclear magnetism (Oxford University Press, London, 1961).
        \bibitem{ZhengNuQ}G.-q. Zheng $et$ $al$., J. Phys. Soc. Jpn. {\bf 64}, 2524 (1995).
        \bibitem{SrPt2As2}S. Kawasaki $et$ $al$., Phys. Rev. B {\bf 91}, 060510(R) (2015).
        \bibitem{Narath}
        A. Narath, Phys. Rev. {\bf 162}, 320 (1967).
        
        
        
        \bibitem{MoriyaJPSJ}
        T. Moriya,  J. Phys. Soc. Jpn. {\bf 18}, 516 (1963). 
        \bibitem{Obata} Y. Obata, J. Phys. Soc. Jpn. {\bf 19}, 2348 (1964).
        \bibitem{AWHunt}A. W. Hunt $et$ $al$., Phys. Rev. B {\bf 64}, 134525 (2001).
        \bibitem{Mei}J.-W. Mei $et$ $al$., Phys. Rev. B {\bf 85}, 134519 (2012).
        
        \bibitem{Blinc}R. Blinc and T. Apih, Prog. Nucl. Magn. Reson. Spectrosc. {\bf 41}, 49, (2002).
        
        \bibitem{CominNatMat}R. Comin $et$ $al$., Nature Mater. {\bf 14}, 796 (2015).
        \bibitem{DavisPNAS}K. Fujita $et$ $al$., PNAS {\bf 111}, E3026 (2014).
        
        \bibitem{JulienPRL}
        J. Ka\v{c}mar\v{c}\'{i}k $et$ $al$., Phys. Rev. Lett. {\bf 121}, 167002 (2018).
        
        
        
        \bibitem{Hsieh}L. Zhao $et$ $al$., Nature Phys. {\bf 13}, 250 (2017).
        
        
        
        
        \bibitem{KitaokaT1} 
        Y. Kitaoka $et$ $al$., J. Phys. Soc. Jpn. {\bf 57}, 30 (1988).
        
        \bibitem{ImaiT1}
        T. Imai $et$ $al$., J. Phys. Soc. Jpn. {\bf 57}, 1771 (1988).
        
        \bibitem{Ishida}
        K. Ishida $et$ $al$., J. Phys. Soc. Jpn. {\bf 89}, 064707 (2020). 
        
        \bibitem{Girod}
        C. Girod $et$ $al$., Phys. Rev. B {\bf 103}, 214506 (2021).
        
        \bibitem{Lizaire}
        M. Lizaire $et$ $al$., Phys. Rev. B {\bf 104}, 014515 (2021).
           
        
        \bibitem{Matsuda}
        Y. Sato $et$ $al$., Nature Phys. {\bf 13}, 1074 (2017).
                    
        
        
        
        
    \end{thebibliography}
\end{document}